\documentstyle[aps,prl,epsf,epsfig]{revtex}

\begin{document}
\def\CC{{\rm\kern.24em \vrule width.04em height1.46ex depth-.07ex
\kern-.30em C}}
\def\P{{\rm I\kern-.25em P}}
\def\RR{{\rm
         \vrule width.04em height1.58ex depth-.0ex
         \kern-.04em R}}

\newcommand{\bq}{\begin{eqnarray}}
\newcommand{\eq}{\end{eqnarray}}
\newcommand{\no}{\nonumber\\}

\draft
\title{
Geometrical phases for the $G(4,2)$ Grassmannian manifold }
\author{
Regina Karle$^{1}$ and Jiannis Pachos$^{2,3}$ }
\address{
$^1$ Ludwig-Maximilians-Universit\"at, M\"unchen, Germany\\ $^2$Max-Planck-Institut
f\"ur Quantenoptik, D-85748 Garching, Germany \\ $^3$Optics Section, Blackett
Laboratory, Imperial College London, London SW7 2BZ, U.K.
\footnote{Present address. Email: jiannis.pachos@imperial.ac.uk}
}
\date{\today}
\maketitle
\begin{abstract}
We generalize the usual abelian Berry phase generated for example in a system with
two non-degenerate states to the case of a system with two doubly degenerate energy
eigenspaces. The parametric manifold describing the space of states of the first
case is formally given by the $G(2,1)$ Grassmannian manifold, while for the
generalized system it is given by the $G(4,2)$ one. For the latter manifold which
exhibits a much richer structure than its abelian counterpart we calculate the
connection components, the field strength and the associated geometrical phases that
evolve non-trivially both of the degenerate eigenspaces. A simple atomic model is
proposed for their physical implementation.
\end{abstract}


\section{Introduction}

Geometrical phases have attracted much interest since the seminal work by Berry
\cite{Berry}. The simple example of the abelian Berry phase produced, e.g. by the
adiabatic transition of a spin-1/2 particle that follows a rotating magnetic field,
has found many applications in quantum optics, molecular physics and so on.
Theoretically, it was extended to non-abelian phases by Wilczek and Zee \cite{WIZE}
by additionally employing to the setup of the usual Berry phase a degenerate
structure that allows geometrical unitary evolutions of higher dimensionality
describing transitions of population within the degenerate eigenspace. There are
different applications of geometrical evolutions in the literature \cite{Fujii} and
in particular related to quantum information. Special attention has been given to
the evolutions that are parameterized by the $n$-dimensional complex projective
manifold, $CP^n$. It corresponds to the parametric manifold ${\cal M}$ of a physical
model where $n$ states out of $n+1$ are preserved degenerate at all times
\cite{NAK}. For this case the parametric space ${\cal M}$ is given by
\begin{eqnarray}
CP^n \approx { U(n+1) \over U(n) \times U(1)}\,\, ,
\nonumber
\end{eqnarray}
where the dictated control transformations are between the non-degenerate state and
each degenerate one. By performing adiabatically closed paths in this parametric
space geometrical unitary operators are generated, called holonomies, $\Gamma$,
acting solely on the degenerate states. Their relevance for performing quantum
computation was first demonstrated in \cite{PAZARA}. Since then they enjoyed
theoretical applications in quantum optical models with laser beams \cite{Pachos3},
trapped ions \cite{Duan}, optical cavities \cite{Pachos1} or quantum dots
\cite{Solinas}.

A further generalization of the parametric control space is realized by employing
the Grassmannian manifolds. They are given by the coset space structure
\begin{eqnarray}
G(m,n) \approx { U(m) \over U(n) \times U(m-n)} \,\, .
\nonumber
\end{eqnarray}
The complex projective manifolds can be considered as a special case of the
Grassmannian ones. In particular, for $m=2$ and $n=1$ we obtain the $CP^1$ space
where the Berry phases are defined, while for $m=n+1$ we have the identity $G(n+1,n)
\approx CP^{n}$. There are two degenerate eigenspaces corresponding to the $G(m,n)$
model that are $n$ and $m-n$ dimensional. With the same adiabatic control procedure
in the parametric space ${\cal M}=G(m,n)$ non-trivial holonomies are generated in
both of them, contrary to all previously studied examples. Here, we shall restrict
to the $G(4,2)$ case, where the connection components, the corresponding field
strengths and a set of holonomies will be explicitly given.

\section{Holonomies and Physical models}

\subsection{Berry phases}

\begin{center}
\begin{figure}[t]
\centerline{
\epsffile{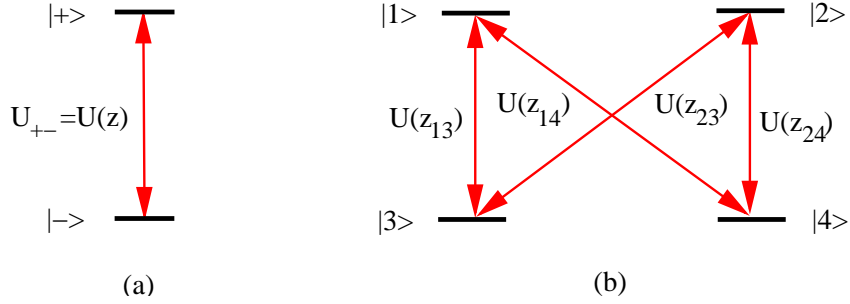}}
\vspace{0.5cm}
\caption[contour]{\label{system} Schematic state-structure and interactions
parameterized by (a) the $G(2,1)$ and for (b) the $G(4,2)$ Grassmannian manifolds.
The arrows represent $U(2)$ transformations between the connected states implemented
e.g., by Raman transitions, while due to the coset structure there are no
transformations between the degenerate states of (b).}
\end{figure}
\end{center}

The well known Berry phase can be produced by performing loops in the $G(2,1)\approx
CP^1$ parametric space of external control parameters that determine the unitary
evolution of two level system as sketched in Figure \ref{system}(a). Let us present
one possible Berry phase implementation in atomic physics.

Consider the case of an atom with two non-degenerate ground levels $|+\rangle$ and
$|-\rangle$, with corresponding eigenvalues $E_+=\omega/2$ and $E_-=-\omega/2$, and
an excited one $|e\rangle$. By performing a Raman adiabatic transfer \cite{Gulley}
between levels $|+\rangle$, $|-\rangle$ with the help of two detuned laser fields
with Rabi frequencies $\Omega_+$ and $\Omega_-$ and common detuning $\Delta$ it is
possible to adiabatically eliminate level $|e\rangle$ and create in the basis
$\{|+\rangle, |-\rangle\}$ the following unitary transformation
\[
{\cal U}(\theta,\phi)=\left( \begin{array}{cc}
\cos {\theta \over 2}  & i e^{ i\phi} \sin {\theta \over 2}
\\
ie^{-i\phi} \sin {\theta \over 2}
&
\cos {\theta \over 2}
      \end{array} \right) ,
\]
where $\theta= 2 |\Omega_{\text{eff}}| t$, $\phi=\text{arg} \{\Omega_{\text{eff}}\}$
and $\Omega_{\text{eff}}=\Omega_+ \Omega_-^* /(4 \Delta)$. In the above evolution we
have neglected the A.C. Stark shift effect which can be compensated with properly
detuned lasers. In accordance the initial free Hamiltonian of the $|+\rangle$ and
$|-\rangle$ states, given by $H_0={\omega \over 2} \sigma_z$, is transformed as
\begin{eqnarray}
H= {\omega \over 2} {\cal U}(\theta,\phi) \sigma_z {\cal U}^\dagger(\theta,\phi)
\,\, .
\nonumber
\end{eqnarray}
Assume that at time $t=0$ we have $\theta=\phi=0$ and the initial state of the
system is given by $|\Psi_\pm(0)\rangle=|\pm \rangle$. Let us consider the evolution
resulting when a closed path $C$ is spanned very slowly in the parametric plane
$(\theta, \phi)$ in time interval $T$. Due to cyclicity and adiabaticity no
population has moved from one state to the other at the end of the loop, that is,
the energy of the system has not changed. Still the states are allowed to obtain a
geometrical phase factor at the end of this evolution, which can be calculated from
the time evolution operator of the system given by the Schr\"odinger equation. A
dynamical phase does not appear due to the geodesic character of Raman evolutions.
In particular, we obtain that $|\Psi_+(T) \rangle =
\Gamma^{ + }_A(C) |+\rangle$ and $| \Psi_-(T) \rangle =\Gamma^{ -
}_A(C) |-\rangle$ that is the states acquire a geometrical phase factor that depend
on the spanned loop $C$, and a connection $A$. The components of the latter are
defined, and analytically given for this case, by
\begin{eqnarray}
A^\pm_\theta= \langle \pm| {\cal U}^\dagger {\partial \over \partial
\theta}{\cal U} |\pm\rangle =0
\,\,\,\,\,\,\,\,\,\,\,
\text{and}
\,\,\,\,\,\,\,\,\,\,\,
A^\pm_\phi= \langle \pm| {\cal U}^\dagger {\partial \over \partial \phi} {\cal U}
|\pm\rangle= \pm i\sin ^2 \theta \,\, .
\nonumber
\end{eqnarray}
In terms of the connection $A$ the phase factors $\Gamma^{\pm}_A(C)\equiv e^{i
\varphi_{\pm}}$ are given by
\begin{equation}
\Gamma^{\pm}_A(C)=\exp{\oint_C A^\pm } = \exp \int \!\!\! \int_{\Sigma(C)}
d\theta d \phi \,\, F^\pm_{\theta \phi}(\theta,\phi) \,\, ,
\label{Stokes}
\end{equation}
where $F^\pm_{\theta \phi}(\theta,\phi)=\partial_\theta A^\pm _\phi -\partial_\phi
A^\pm _\theta=\pm i\sin 2 \theta$ is the non-zero component of the field strength
associated with $A$ and $\Sigma(C)$ is the surface on the plane $(\theta, \phi)$
bounded by the loop $C$. The second step in (\ref{Stokes}) is due to Stokes theorem.
For any $C$ we obtain the relation $\varphi_+=-\varphi_-=\int \!\!\!
\int_{\Sigma(C)}d\theta d \phi \,\, \sin 2 \theta$ between the two
Berry phases. Note that $\varphi_+=\Omega/2$ where $\Omega$ is the solid angle
spanned by the circulation of a unit vector with directions given by the
$(\theta,\phi)$ angles. The half factor reflects the spin-1/2 transformation
properties of the employed two atomic levels. This physical evolution produces a
measurable abelian Berry phase. Their generalization to unitary matrices
(holonomies) with the employment of degenerate structures is given in the following
for the case of the Grassmannian manifold $G(4,2)$.

\subsection{Holonomies for the Grassmannian manifold $G(4,2)$}

The Grassmannian manifold $G(4,2)$ corresponds to the parametric space of the
Hamiltonian $H={\cal U} H_0 {\cal U}^\dagger$ where $H_0=\omega/2\, \text{diag}
(1,1,-1,-1)$ and ${\cal U}$ are $U(4)$ unitary transformations that act
non-trivially on the Hamiltonian $H_0$. Clearly, $H_0$ has the following
eigenvectors $|+_1\rangle\equiv|1\rangle= (1,0,0,0)$,
$|+_2\rangle\equiv|2\rangle=(0,1,0,0)$, $|-_1\rangle\equiv |3\rangle=(0,0,1,0)$ and
$|-_2\rangle\equiv |4\rangle =(0,0,0,1)$. The first two states span the degenerate
eigenspace $S_+$ with eigenvalue $E_+=\omega/2$, while the last two span the
eigenspace $S_-$ with eigenvalue $E_-=-\omega/2$. The schematic representation of
this model is given in Figure
\ref{system}(b), where the eligible transformations are depicted by arrows. Each
arrow correspond to a $U(2)$ transformation $U(z_{ij})$ parameterized by a complex
number $z_{ij}=\theta_{ij} \exp i \phi_{ij}$ for $i =1,2$ and $j=3,4$. Hence, the
real decomposition of the parametric space $G(4,2)$ is given by ${\cal M}\equiv
\{\theta_{ij}, \phi_{ij}\}\equiv\{\sigma_{ij}\}$. Here we adopt the following
ordering for the general unitary transformation, ${\cal
U}(\sigma)=U(z_{13})U(z_{14})U(z_{23}) U(z_{24})$.

Now we can define the connection components for each degenerate
eigenspace. They are given from the following equation
\begin{equation}
(A^{\pm}_{\sigma})_{\alpha \beta}\equiv \langle \alpha | {\cal U}^\dagger (\sigma)
{\partial \over \partial \sigma} {\cal U} (\sigma) |\beta \rangle \,\, ,
\label{connection}
\end{equation}
where the basis vector $|\alpha\rangle$ and $|\beta\rangle$ belong in the same
degenerate eigenspace of $H_0$. From (\ref{connection}) we see that the matrix
$A^{\pm}_{\sigma}$ is anti-hermitian. For $\{\alpha,\beta\} =\{1,2\}$ we obtain the
elements of the $2\times 2$ matrix $A^+$, while for $\{\alpha,\beta\}=\{3,4\}$ we
obtain the elements of $A^-$. The holonomic unitary operator, generated when a
closed path, $C$, is spanned adiabatically in the space of the control parameters
${\cal M}$, is defined by
\begin{eqnarray}
\Gamma^{\pm}_A(C) \equiv {\bf P} \exp \oint_C A^\pm \,\, .
\nonumber
\end{eqnarray}
The path ordering symbol does not allow to calculate the path integral first and
then to exponentiate the result as the different components of the connection $A$ do
not commute in general. For those cases the simple form of Stokes theorem does not
apply \cite{Karp}. Still it is possible to analytically calculate the holonomies
which result from the commuting components of the connection $A$ in the same way as
we did for the abelian case. Alternatively, the non-abelian version of Stokes
theorem can be employed as in \cite{Pachos3,Pachos2}.

Before moving to the analytic calculation of the holonomies, let us see how the
transformations ${\cal U} (\sigma)$ can be physically realized for example by an
atomic system. According to Figure \ref{system}(b) we can generalize the model of
the two level atom with four levels, pair-wise degenerate and apply laser beams
connecting these states via additional exciting states by Raman adiabatic transfers.
As in the case of the two-level system presented in the previous subsection, a
$U(2)$ unitary transformation results from each Raman transition. Successive
applications of those unitaries are able to construct the general unitary
transformation ${\cal U}(\sigma)$ parameterized by the $G(4,2)$ manifold. A detailed
study of the generation of the Berry phases with ${\cal M}=CP^2$ control manifold
with an atomic system manipulated by Raman transitions is given in \cite{Unanyan}.

\section{Connection and Field Strength Components}

\subsection{Connection $A$}

In this subsection we shall present the connection components related to the
parametric space $G(4,2)$ by employing definition (\ref{connection}). In particular,
they are $2 \times 2$ matrices paired in the following with respect to the $S_+$ and
$S_-$ degenerate eigenspaces they act on. Thus, we obtain

\[
A^+ _{\theta_{13}}=\sin \theta_{23}\cos \theta_{14}\cos \theta_{24} \left(
\begin{array}{cc} 0 & - e^{ i(\phi_{13}-\phi_{23})}
\\
e^{-i(\phi_{13}-\phi_{23})}
&
2 i \sin (\phi_{13}-\phi_{14}+\phi_{24}-\phi_{23}) \tan \theta_{14} \sin \theta_{24}
      \end{array} \right) ,
\]

\[
A^- _{\theta_{13}}=\sin \theta_{14} \cos \theta_{23} \cos \theta_{24}
\left( \begin{array}{cc}
0
&
- e^{ i(\phi_{14}-\phi_{13})}
\\
e^{-i(\phi_{14}-\phi_{13})}
&
2 i \sin (\phi_{14}+\phi_{23}-\phi_{13}-\phi_{24})
\tan \theta_{23} \sin \theta_{24}
      \end{array} \right) ,
\]

\[
A^+ _{\theta_{14}}=\sin \theta_{24}
\left( \begin{array}{cc}
0
&
-e^{i(\phi_{14}-\phi_{24})}
\\
e^{-i(\phi_{14}-\phi_{24})}
& 0
      \end{array} \right) ,
\,\,\,\,\,\,\,\,\,\,\,\,\,\,\,\,\,\,
A^- _{\theta_{14}}=\left( \begin{array}{cc} 0 & 0
\\
0 & 0
      \end{array} \right) ,
\]

\[
A^+ _{\theta_{23}}=\left( \begin{array}{cc}
0 & 0
\\
0 & 0
      \end{array} \right) ,
\,\,\,\,\,\,\,\,\,\,\,\,\,\,\,\,\,\,
A^- _{\theta_{23}}=\sin \theta_{24}
\left( \begin{array}{cc}
0 & -e^{i(\phi_{24}-\phi_{23})}
\\
e^{-i(\phi_{24}-\phi_{23})} & 0
      \end{array} \right) ,
\]

\[
A^+ _{\theta_{24}}=\left( \begin{array}{cc}
0 & 0
\\
0 & 0
      \end{array} \right) ,
\,\,\,\,\,\,\,\,\,\,\,\,\,\,\,\,\,\,
A^- _{\theta_{24}}=\left( \begin{array}{cc}
0 & 0
\\
0 & 0
      \end{array} \right) .
\]

The matrix elements of $A _{\phi_{13}}$ are given by

\bq
&&
(A^+ _{\phi_{13}})_{11}=-i \sin^2 \theta_{13} \cos^2 \theta_{14} \,\, ,
\no \no
&&
(A^+ _{\phi_{13}})_{12}=-\frac{1}{2} i e^{i(\phi_{13}-\phi_{23})}\sin
2 \theta_{13} \sin \theta_{23} \cos \theta_{14} \cos
\theta_{24}+\frac{1}{2} i e^{-i(\phi_{14}-\phi_{24})} \sin^2
\theta_{13} \sin 2 \theta_{14} \sin \theta_{24} \,\, ,
\no\no
&&
(A^+ _{\phi_{13}})_{21}=(A^+ _{\phi_{13}})_{12}^* \,\, ,
\no\no
&&
(A^+ _{\phi_{13}})_{22}=
i \sin^2 \theta_{13} \sin^2 \theta_{23} \cos^2 \theta_{24}
-i \sin^2 \theta_{13} \sin^2 \theta_{24} \sin^2 \theta_{14}
\no \no
&&
\,\,\,\,\,\,\,\,\,\,\,\,\,\,\,\,\,\,\,\,\,\,\,\,\,\,\,\,\,
+\frac{1}{2} i \cos (\phi_{23}-\phi_{13}+\phi_{14}-\phi_{24}) \sin 2 \theta_{13} \sin
\theta_{14} \sin \theta_{23} \sin 2 \theta_{24}\,\, ,
\nonumber
\eq

\bq
&&
(A^- _{\phi_{13}})_{11}=i \sin^2 \theta_{13} \cos^2 \theta_{23} \,\, ,
\no \no
&&
(A^- _{\phi_{13}})_{12}=
-\frac{1}{2} i e^{-i(\phi_{23}-\phi_{24})}
\sin^2 \theta_{13} \sin 2 \theta_{23} \sin \theta_{24}+\frac{1}{2}
i e^{-i(\phi_{13}-\phi_{14})} \sin 2 \theta_{13} \sin \theta_{14}
\cos \theta_{23} \cos \theta_{24}\,\, ,
\no \no
&&
(A^- _{\phi_{13}})_{21}=(A^- _{\phi_{13}})_{12}^* \,\, ,
\no \no
&&
(A^- _{\phi_{13}})_{22}=
i \sin^2 \theta_{13} \sin^2
\theta_{23} \sin^2 \theta_{24}-i \sin^2 \theta_{13} \sin^2
\theta_{14} \cos^2 \theta_{24}
\no\no
&&
\,\,\,\,\,\,\,\,\,\,\,\,\,\,\,\,\,\,\,\,\,\,\,\,\,\,\,\,\,
-\frac{1}{2} i \cos
(-\phi_{14}+\phi_{13}+\phi_{24}-\phi_{23}) \sin 2 \theta_{13} \sin
\theta_{14} \sin \theta_{23} \sin 2 \theta_{24}\,\, ,
\nonumber
\eq

\[
A^+ _{\phi_{14}}=i \sin^2 \theta_{14}
\left( \begin{array}{cc}
-1
&
-e^{i(\phi_{14}-\phi_{24})} \text{ctan} \, \theta_{14} \sin \theta_{24}
\\
-e^{-i(\phi_{14}-\phi_{24})} \text{ctan} \, \theta_{14} \sin \theta_{24}
&
\sin^2 \theta_{24}
      \end{array} \right) ,
\]

\[
A^- _{\phi_{14}}=i \sin^2 \theta_{14} \cos^2 \theta_{24}
\left( \begin{array}{cc}
0 & 0
\\
0 &  1
      \end{array} \right) ,
\]

\[
A^+ _{\phi_{23}}=i \sin^2 \theta_{23} \cos^2 \theta_{24}
\left( \begin{array}{cc}
0 & 0
\\
0 & -1
      \end{array} \right) ,
\]

\[
A^- _{\phi_{23}}=i \sin^2 \theta_{23}
\left( \begin{array}{cc}
1
&
e^{i(\phi_{24}-\phi_{23})} \text{ctan} \, \theta_{23} \sin \theta_{24}
\\
e^{-i(\phi_{24}-\phi_{23})} \text{ctan} \, \theta_{23} \sin \theta_{24}
&
-\sin^2 \theta_{24}
      \end{array} \right) ,
\]

\[
A^+ _{\phi_{24}}=i \sin^2 \theta_{24}
\left( \begin{array}{cc}
0 & 0
\\
0 & - 1
      \end{array} \right) ,
\,\,\,\,\,\,\,\,\,\,\,\,\,\,\,\,\,\,
A^- _{\phi_{24}}=i \sin^2 \theta_{24}
\left( \begin{array}{cc}
0 & 0
\\
0 &  1
      \end{array} \right) .
\]

\subsection{Field strength $F$}

In this subsection we shall calculate the field strength $F$ associated with the
previous connections $A$. Its components are given by $F_{\mu \nu}=\partial_\mu
A_\nu - \partial_\nu A_\mu +[A_\mu, A_\nu]$. Here, we shall restrict on the field
strength components for which the commutator part $[A_\mu,A_\nu]$ is zero for all
values of the parameters $\theta_{ij}$ and $\phi_{ij}$. For them the computation of
their holonomies is straightforward as we shall see in the following subsection. Let
us present first the $(\theta,\phi)$ field strength components. The ones that are
obtained from commuting connection components acting on the $S_+$ and $S_-$
eigenspaces are given by

\bq
&&
(F^+ _{\theta_{24} \phi_{13}})_{11}=0 \,\, ,
\no \no
&&
(F^+ _{\theta_{24} \phi_{13}})_{12}=
{i \over 2} e^{i(\phi_{13} - \phi_{23})} \sin 2 \theta _{13}
\sin \theta _{23} \cos \theta _{14} \sin \theta _{24} +
{i \over 2} e^{-i(\phi_{14} - \phi_{24})} \sin 2 \theta _{14}
\sin ^2 \theta _{13} \cos \theta _{24} \,\, ,
\no \no
&&
(F^+ _{\theta_{24} \phi_{13}})_{21}= (F^+ _{\theta_{24}
\phi_{13}})_{12}^*\,\, ,
\no \no
&&
(F^+ _{\theta_{24} \phi_{13}})_{22}=
-i \sin \theta _{13} \sin 2 \theta_{24} (\sin^2 \theta_{14} + \sin ^2
\theta_{23})
+
i \cos ( \phi_{23} - \phi_{13} + \phi_{14} - \phi_{24}) \sin 2
\theta_{13} \sin \theta_{14} \sin \theta_{23} \cos 2 \theta_{24}\,\, ,
\nonumber
\eq

\bq
&&
(F^- _{\theta_{24} \phi_{13}})_{11}=0 \,\, ,
\no \no
&&
(F^- _{\theta_{24} \phi_{13}})_{12}=
- {i \over 2} e^{i(\phi_{23} -\phi_{24})} \sin^2 \theta_{13} \sin 2
\theta_{23} \cos \theta_{24}
-{i \over 2} e^{i(\phi_{13} -\phi_{14})} \sin 2 \theta_{13} \sin
\theta_{14} \cos \theta_{23} \sin \theta_{24}\,\, ,
\no \no
&&
(F^- _{\theta_{24} \phi_{13}})_{21}=(F^- _{\theta_{24}
\phi_{13}})_{12}^* \,\, ,
\no \no
&&
(F^- _{\theta_{24} \phi_{13}})_{22}=
i \sin ^2 \theta_{13} \sin 2 \theta_{24} (\sin ^2 \theta_{23} + \sin^2
\theta _{14}) - i \cos (\phi_{13} - \phi _{14} + \phi _{24} - \phi
_{23}) \sin 2 \theta_{13} \sin \theta_{14} \sin \theta_{23} \cos 2
\theta_{24} \,\, ,
\nonumber
\eq

\[
F^+ _{\theta_{24} \phi_{14}}=\left( \begin{array}{cc} 0 &
-\frac{1}{2} i e^{i(\phi_{14}-\phi_{24})} \sin 2 \theta_{14} \cos
\theta_{24}
\\
-\frac{1}{2} i e^{-i(\phi_{14}-\phi_{24})} \sin 2 \theta_{14} \cos
\theta_{24} & i \sin^2 \theta_{14} \sin 2 \theta_{24}
      \end{array} \right) ,
\]

\[
F^- _{\theta_{24} \phi_{14}}=
-i \sin^2 \theta_{14} \sin 2 \theta_{24}
\left( \begin{array}{cc}
0 & 0
\\
0 & 1
      \end{array} \right) ,
\]

\[
F^+ _{\theta_{24} \phi_{23}}=
i \sin^2 \theta_{23} \sin 2 \theta_{24}
\left( \begin{array}{cc}
0 & 0
\\
0 & 1
      \end{array} \right) ,
\]

\[
F^- _{\theta_{24} \phi_{23}}=\left( \begin{array}{cc} 0 &
\frac{1}{2} i e^{i(\phi_{24}-\phi_{23})} \sin 2 \theta_{23} \cos
\theta_{24}
\\
\frac{1}{2} i e^{i(-\phi_{24}+\phi_{23})} \sin 2 \theta_{23} \cos
\theta_{24} & -i \sin^2 \theta_{23} \sin 2 \theta_{24}
      \end{array} \right) ,
\]

\[
F^+ _{\theta_{24} \phi_{24}}=
-i \sin 2 \theta_{24}
\left( \begin{array}{cc}
0 & 0
\\
0 & 1
      \end{array} \right) ,
\,\,\,\,\,\,\,\,\,\,\,\,\,\,\,\,\,\,
F^- _{\theta_{24} \phi_{24}}=
i \sin 2 \theta_{24}
\left( \begin{array}{cc}
0 & 0
\\
0 & 1
      \end{array} \right) ,
\]

From the connection components where only the ones that act on $S_+$ commute we
obtain the following field strength components

\bq
F^+ _{\theta_{23} \phi_{13}}= &&
{i \over 2} \cos \theta_{24} \cos \theta_{23} \sin 2 \theta_{13}
\no \no
&&
\times \left( \begin{array}{cc}
0
&
-e^{i(\phi_{13}-\phi_{23})}  \cos \theta_{14}
\\
-e^{-i(\phi_{13}-\phi_{23})} \cos \theta_{14}
&
2 \tan \theta_{13} \sin  \theta _{23} +
2 \cos (\phi_{23} -\phi_{13} + \phi_{14} -\phi_{24})
\sin \theta_{14} \sin \theta_{24}
      \end{array} \right) ,
\eq

\[
F^+ _{\theta_{23} \phi_{23}}=
-i \cos^2 \theta_{24} \sin 2 \theta_{23}
\left( \begin{array}{cc}
0 & 0
\\
0 & 1
      \end{array} \right) ,
\]
while $F^+_{\theta_{23} \phi_{14}}=F^+_{\theta_{23} \phi_{24}}=0$. From the
connection components where only the ones that act on $S_-$ commute we obtain the
following field strength components

\bq
F^- _{\theta_{14} \phi_{13}}=&&
-{ i \over 2 } \sin 2 \theta _{13} \cos \theta_{14} \cos \theta_{24}
\no \no
&&
\times\left( \begin{array}{cc}
0
&
- e^{-i(\phi_{13}-\phi_{14})} \cos \theta_{23}
\\
- e^{ i(\phi_{13}-\phi_{14})} \cos \theta_{23}
&
2 \tan \theta_{13} \sin \theta_{14} \cos \theta_{24} +
2 \cos (\phi_{13} -\phi_{14} + \phi_{24} -\phi_{23})
\sin \theta_{23} \sin \theta_{24}
      \end{array} \right) ,
\eq

\[
F^- _{\theta_{14} \phi_{14}}=
i \cos^2 \theta_{24} \sin 2 \theta_{14}
\left( \begin{array}{cc}
0 & 0
\\
0 & 1
      \end{array} \right) ,
\]

while $F^-_{\theta_{14} \phi_{23}}=F^-_{\theta_{14} \phi_{24}}=0$. Finally, the
$(\theta,\theta)$ field strength components are given by

\[
F^+ _{\theta_{13} \theta_{24}}=
\sin \theta_{23} \cos \theta_{14}
\left( \begin{array}{cc}
0 & - e^{i(\phi_{13} - \phi_{23})} \sin \theta_{24}
\\
e^{-i(\phi_{13} - \phi_{23})} \sin \theta_{24}  & -2i \sin (\phi_{13} -
\phi_{14}+ \phi_{24} -\phi_{23}) \tan \theta_{14} \cos 2 \theta _{24}
      \end{array} \right) ,
\]

\[
F^- _{\theta_{13} \theta_{24}}=
\sin \theta_{14} \cos \theta_{23}
\left( \begin{array}{cc}
0 & - e^{i(\phi_{14} - \phi_{13})} \sin \theta_{24}
\\
e^{-i(\phi_{14} - \phi_{13})} \sin \theta_{24}  & -2i \sin (\phi_{14} +
\phi_{23}- \phi_{13} -\phi_{24}) \tan \theta_{23} \cos 2 \theta _{24}
      \end{array} \right) ,
\]

\begin{eqnarray}
F^+_{\theta_{14} \theta_{23}}= F^-_{\theta_{14} \theta_{23}}= 0
\nonumber
\end{eqnarray}

\[
F^+ _{\theta_{14} \theta_{24}}=
- \cos \theta_{24}
\left( \begin{array}{cc}
0 & - e^{i(\phi_{14} - \phi_{24})}
\\
e^{-i(\phi_{14} - \phi_{24})} & 0
      \end{array} \right) ,
\,\,\,\,\,\,\,\,\,\,\,\,\,\,\,\,\,\,
F^-_{\theta_{14} \theta_{13}}= 0
\]

\[
F^+_{\theta_{23} \theta_{24}}= 0
\,\,\,\,\,\,\,\,\,\,\,\,\,\,\,\,\,\,
F^- _{\theta_{23} \theta_{24}}=
- \cos \theta_{24}
\left( \begin{array}{cc}
0 & - e^{i(\phi_{24} - \phi_{23})}
\\
e^{-i(\phi_{24} - \phi_{23})} & 0
      \end{array} \right) ,
\]

\[
F^+ _{\theta_{13} \theta_{23}}=
-\cos \theta_{23}\cos \theta_{14}\cos \theta_{24}
\left( \begin{array}{cc}
0 & - e^{ i(\phi_{13}-\phi_{23})}
\\
e^{-i(\phi_{13}-\phi_{23})}
&
2 i \sin (\phi_{13}-\phi_{14}+\phi_{24}-\phi_{23}) \tan \theta_{14} \sin \theta_{24}
      \end{array} \right) ,
\]

\[
F^- _{\theta_{13} \theta_{14}}=
-\cos \theta_{14} \cos \theta_{23} \cos \theta_{24}
\left( \begin{array}{cc}
0
&
- e^{ i(\phi_{14}-\phi_{13})}
\\
e^{-i(\phi_{14}-\phi_{13})}
&
2 i \sin (\phi_{14}+\phi_{23}-\phi_{13}-\phi_{24})
\tan \theta_{23} \sin \theta_{24}
      \end{array} \right) .
\]

\subsection{Holonomies $\Gamma$}

In general, the explicit calculation of the Holonomies of matrix connections $A$ is
non-straightforward as they involve the path ordering procedure when exponentiating
their loop integral. On the other hand it is possible to restrict our cyclic
evolutions to specific planes $(\sigma,\sigma')$ that correspond to commuting
components $A_\sigma$ and $A_{\sigma'}$. For those loops we can employ Stokes
theorem for the abelian theories and obtain
\begin{equation}
\Gamma^{\pm}_A(C) \equiv {\bf P} \exp \oint_C A^\pm = \exp \oint_C A^\pm
= \exp \int \!\!\! \int_{\Sigma(C)}
d\sigma d \sigma' \,\, F^\pm_{\sigma \sigma'}(\sigma,\sigma') \,\, ,
\label{holonomy}
\end{equation}
where the rest of the variables are considered constant. The path ordering symbol
has been taken out at the second step as the connection components on the
$(\sigma,\sigma')$ commute with each other. Hence, Stokes theorem can be applied
straightforwardly as presented in the previous section for the abelian Berry phase.
An analytic expression for the holonomies can be obtained by exponentiating the $2
\times 2$ matrices resulting from the surface integral of the field strength
components presented in the previous subsection.

\section{Discussion}

A theoretical model has been presented for the construction of non-abelian
holonomies for the $G(4,2)$ Grassmannian manifold. This is the generalization of the
usual abelian Berry phase to the case of a quantum system consisting of a doubly
degenerate energy eigenspace. The evolution of both degenerate spaces has been
presented that are produced from the same cyclic adiabatic evolution. The main
difference with the abelian case is that it is possible to have manipulations of
state population in each degenerate space rather than just generation of overall
phase factors. This can be achieved for example by spanning loops $C$ on the
$(\theta_{24}, \phi_{13})$ plane where population can be interchanged in a well
defined way between the states $|1\rangle$ and $|2\rangle$ as well as between
$|3\rangle$ and $|4\rangle$. In contrast, loops on the $(\theta_{24}, \phi_{24})$
plane contribute only Berry-like phases on the conjugate states $|2 \rangle$ and
$|4\rangle$ (see Figure \ref{system}). It is worth noticing that the holonomic
evolution is not producing any correlations between the two different eigenspaces
$S_+$ and $S_-$ due to the cyclicity of the adiabatic procedure. Even though it is
possible to have each degenerate eigenspace evolving with a different holonomy there
is a correspondence between the evolutions as can be easily seen in the previous
section facilitating eventually their detection in a physical system and the
verification of the above results. Indeed, the components $F^+$ and $F^-$ of the
field strength have a similar functional dependence on the variables
$(\theta_{ij},\phi_{ij})$ and their surface integral in (\ref{holonomy}) bears a
common dependence in the area $\Sigma(C)$. These holonomic characteristics can be
verified in the laboratory with present technology by employing an atomic cloud and
manipulating the atomic states with external laser beams.

\end{document}